\documentclass[preprint]{elsarticle}
\usepackage[utf8]{inputenc}
\usepackage[english]{babel}
\usepackage{amssymb}
\usepackage{amsmath}
\usepackage{multirow}
\usepackage{textcomp}
\usepackage{gensymb}

\usepackage{lineno}

\usepackage{setspace}
\begin{document}
\begin{frontmatter}

\title{Stability criterion for fresh cement foams}

\author[navier,uio]{Blandine Feneuil\corref{cor1}}
\ead{bffeneui@math.uio.no}

\author[navier]{Patrick Aimedieu}

\author[soleil]{Mario Scheel}

\author[soleil]{Jonathan Perrin}

\author[navier]{Nicolas Roussel}

\author[navier]{Olivier Pitois}

\cortext[cor1]{Corresponding author}

\address[navier]{Laboratoire Navier, UMR 8205, École des Ponts ParisTech, IFSTTAR, CNRS, UPE, Champs-sur-Marne, France}

\address[soleil]{Anatomix Beamline, Synchrotron SOLEIL, Saint Aubin, France}

\address[uio]{Current address: Department of Mathematics, University of Oslo, Oslo, Norway}

\begin{abstract}

We prepare and study cement foam samples with well-controlled structure, i.e. containing monodisperse bubbles. We observe that the foam structure often changes before cement setting and identify ripening as the major destabilization mechanism at stake. Drainage plays only a minor role in cement foam destabilization except when bubble size is large. Then we show that a single stability criterion can be defined, for a large range of cement foams with different formulations. This criterion involves the bubble radius and the yield stress of the cement paste such as confined by and between the bubbles, at a given characteristic time after sample preparation.

\end{abstract}


\end{frontmatter}

\section*{Notations}
\begin{small}
\begin{tabular}{p{1.2cm} p{9cm}}
$R_0$ & Initial bubble radius in a cement foam, i.e. bubble radius in the precursor foam\\
$\gamma$ & Air-liquid surface tension\\
$\rho$ & Cement paste density\\
$\rho_{liq}$ & Liquid density (1.0 g/cm$^3$)\\
$\rho_{c}$ & Cement density (3.2 g/cm$^3$)\\
$\Phi$ & Air volume content\\
$\Phi_p$ & In a cement paste, solid content of the paste which depends on the water-to-cement ratio W/C : $\Phi_p=1/(1+\rho_c/\rho_{liq}  W/C)$\\
$W/C_i$ & Water to cement ratio in cement paste before mixing with precursor aqueous foam\\
$W/C_f$ & Water to cement ratio in reference cement paste, i.e. the unfoamed cement paste with the same water and additives content as the cement foam\\
$\tau_{y,0}$ & Yield stress of reference cement paste. This value is measured by spread test.\\
$\tau_{y,foam}$ & Cement foam yield stress measured by start-of-flow experiment\\
$\tau_{y,aq}(\Phi)$ & Yield stress of aqueous foam calculated at air volume content $\Phi$\\
$\tau_{y,int}$ & Yield stress of interstitial cement paste deduced from the cement foam yield stress $\tau_{y,foam}$\\
\end{tabular}
\end{small}

\section{Introduction}

When it is unconstrained, a bubble has a spherical shape because of the air-liquid surface tension. Though, in a foam, bubbles are deformed by their neighbors. The structure of a foam was studied in 1873 by Joseph Plateau, who stated three laws known as Plateau's laws~\cite{2013_Cantat}: (1) two bubbles are separated by a soap film of constant average curvature, (2) three films join in a channel, called Plateau border, forming 120\degree~angles, (3) four Plateau borders join into a node at angle 109.5\degree. The resulting morphology of foam tends to evolve with time due to downward flow of the interstitial fluid due to gravity (drainage), air exchange between bubbles (ripening) and film breakage (coalescence).

\bigbreak

In cement foams, these destabilization mechanisms are expected to occur until cement paste hardening. Thus, to control the final bubble size and air distribution in the foam, one has to stop or slow down the three mechanisms. These mechanisms are affected by initial bubble size. For instance, the increase of bubble size for a given gas volume fraction results in the increase of the size of the film areas between the bubbles, which enhances coalescence. It also increases the size of the Plateau borders and nodes, which favors drainage~\cite{2013_Cantat}. Ripening, on the contrary, is reduced when bubble size increases. Indeed, it is caused by the capillary pressure inside the bubbles $P_c \approx 2 \gamma/R_0$, where $\gamma$ is the air-liquid surface tension and $R_0$ the bubble radius.

To avoid coalescence, liquid film must be stabilized by molecules or partially hydrophobic particles, which adsorb at air-water interfaces. The molecules, called surfactants, must be compatible with the highly alkaline cement solution and be present in sufficiently high amount~\cite{2013_Cantat}.

Consistency of cement paste is also expected to play a major role in foam stability. High yield stress can stop drainage and ripening~\cite{2010_Guignot,2014_Lesov}. However, we have shown in a previous paper~\cite{2019_Feneuil} that, in a cement foam, the effective yield stress of the cement paste confined between the bubbles, noted $\tau_{y,int}$, can differ significantly from the reference yield stress of the cement paste $\tau_{y,0}$, measured in the bubble-free paste. On the one hand, when $\tau_{y,0}$ is low, i.e. a few Pascals, cement grains remain stuck in the channels and nodes between the bubbles, whereas gravity makes the liquid flow to the bottom of the foam. This drainage of the liquid leads to a decrease of the water-to-cement ratio of the interstitial cement paste, and therefore, to an increase of the interstitial yield stress $\tau_{y,int}$ up to about 100 Pa. On the other hand, when the yield stress of the reference cement paste $\tau_{y,0}$ is high, i.e. a few tens of Pascals, no densification of the cement paste through drainage occurs, so that $\tau_{y,int} \approx \tau_{y,0}$ during the first 10 min after sample preparation. For the cement foam formulations studied in~\cite{2019_Feneuil}, cement paste densification through liquid drainage was found to be essential to ensure the foam stability.

Therefore, the stability of fresh cement foams is expected to be observed for pastes with sufficiently high yield stress values, but stability can be observed also for low yield stress values. The aim of this paper is to reconcile those contradictory results and to propose a single stability criterion for cement foams. In the materials and methods, we describe how we prepare cement foams with controlled morphology and formulation, which allows for the factors controlling the stability of these cement foams to be investigated. First, the leading destabilization mechanism is identified. Then, the effects of bubble size and of cement paste yield stress are investigated. Finally, a global criterion for cement foam stability is defined.

\section{Materials and methods}
\subsection{Materials}

\subsubsection{Cement}

We use two cements. The first will be referred to as C1, it is manufactured by Lafarge, in Saint-Vigor factory and C2 is a CEM I cement from Lafarge, Lagerdorf. Their compositions and physical properties are specified in Table~\ref{table_chimie_ciment_c5}.

\begin{table}[!ht]
\begin{center}
\begin{tabular}{|c|c|c|}
\hline
 & C1 & C2\\ \hline
CaO/SiO$_2$ & 3 & 3\\
MgO & 1.1\% & 0.8\%\\
Na$_2$O + 0.658 K$_2$O & 0.34\% & 0.5\%\\
SO$_3$ & 2.58\% & 2.5 \%\\
Cl$^-$ & 0.03\% & 0.04 \%\\
Gypsum & 2.4\% & 4\% \\ \hline
Density (g/cm$^3$) & 3.21 & 3.15\\
SSB (cm$^2$/g) & 3586 & 4330\\ \hline
\end{tabular}
\caption{Chemical and physical properties of cements. C1 refers to CEM I cement from Lafarge, Saint-Vigor and C2 to CEM I cement from Lafarge, Lagerdorf.}
\label{table_chimie_ciment_c5}
\end{center}
\end{table}

\subsubsection{Surfactants}

Two surfactants are used to produce the precursor foam. Tetradecyltrymethyl ammonium bromide (TTAB) is a cationic surfactant at purity above 99\% provided by Sigma-Aldrich. Its molar mass is 336 g/mol. Steol$^\text{\textregistered}$~270 CIT is an anionic surfactant provided by Stepan. Its molar mass indicated by the manufacturer is 382~g/mol and active content  68-72\%. Surfactant chemical formulas can be found in \cite{2017_Feneuil}.

Previous study showed that both surfactants are able to form stable foams in the highly alkaline cement paste interstitial solution~\cite{2017_Feneuil}.

\subsection{Methods}

\subsubsection{Precursor foam and mixing}
\label{part_precursor_foam}

Cement foams are prepared by mixing precursor aqueous foam and cement paste. Precursor foams are generated using the method described in \cite{2019_Feneuil}. Here we recall the main outlines of the method: bubbles are generated with a T-junction (inner diameter $\approx$ 100 $\mu$m). All the bubbles have approximately the same radius $R_0$ (the relative dispersion was measured to be $\Delta R_0 / R_0 \approx 3\%$) which is set by the entrance rates of the gas (nitrogen) and the foaming liquid (water and surfactant). Bubbles are collected in a column, where liquid fraction is controlled thank to addition of foaming liquid from the top of the foam. Studied bubble radii are comprised between 200~$\mu$m and 900~$\mu$m. In TTAB precursor foams, TTAB concentration  is 10~g/L and liquid fraction is between 0.5 and 2\%. In Steol precursor foams, Steol concentration is 1~g/L, and liquid fraction is 1.6$\pm$0.1\% when bubble radius is below 350~$\mu$m and 1.4$\pm$0.1\% otherwise.

Mixing of cement paste and precursor foam is carried out with a flow focusing device as described in \cite{2019_Feneuil}. The main advantage of this method is that bubbles are not broken during the mixing process. Gas volume fraction in the cement foams depends mainly on the flow rates of both the precursor foam and the cement paste. It is however also affected by gas compressibility. By weighting our samples after preparation, we check that air fraction for all of them is between 81\% and 84\% (average value $\Phi=83\%$). Note that the flow focusing method involves the flow of the cement paste in small channels (2~mm diameter) and that therefore, the yield stress of the paste must be limited to a few tens of Pascals to avoid jamming in these channels.

Each sample is obtained by filling a mold (diameter 2.6 cm, height 6 cm), through a layer-by-layer deposition process. Therefore, cement foam does not undergo strong shearing after its preparation, and bubbles are not broken during sample production.

\subsubsection{Protocol}
\label{part_protocol_c5}

\begin{figure}[!ht]
\begin{center}
\includegraphics[width=13cm]{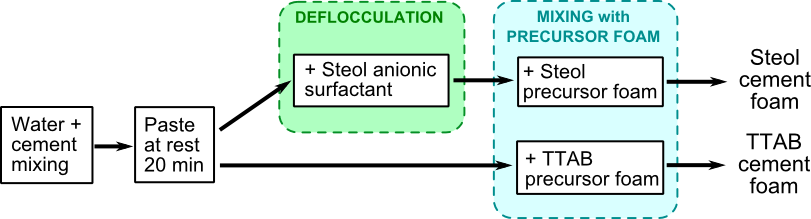}
\caption{Preparation protocol of cement foams made with Steol anionic surfactant and TTAB cationic surfactant.}
\label{schema_protocol_c5}
\end{center}
\end{figure}

For each surfactant, all cement foam samples are prepared using the same mixing procedure, from water and cement mixing to casting. In the case of Steol samples, large amount of surfactant is added to cement paste 20~minutes after cement paste preparation to make the consistency of the paste decrease \cite{2019_Feneuil}. For both surfactants, precursor foam and cement paste are mixed 30~min after cement paste preparation. Our protocol is schematized in Fig.~\ref{schema_protocol_c5}.

\subsubsection{Observation of stability}

Samples are demolded one week after casting and the stability is visually assessed from the final morphology of the cement foam. Sample stability is evaluated according to the scale described below. Illustrations of the scale can be found in \cite{2019_Feneuil}.

\begin{itemize}
\item 3: fully stable sample. All bubbles have kept their original size
\item 2: large stable area. In some areas (oftern the top of the bottom of the sample) the bubble size has changed before cement hardening, but most of the sample is stable
\item 1: small stable area(s). The morphology of most bubbles has evolved, yet, one or several area(s) containing at least several tens of bubbles can be observed.
\item 0: fully unstable sample, no stable area.
\end{itemize}

Moreover the evolution of the bubble morphology of fresh cement foams is followed across the transparent molds. Note however that parietal bubbles are not fully representative of the bulk bubbles. In particular, Plateau borders are larger. For samples with large bubble size or low cement paste yield stress, this leads to the full coverage of the mold walls by cement paste. Obtained images are analyzed with freeware ImageJ program~\cite{2012_Schneider} to calculate the apparent radius of the parietal bubbles. 

\bigbreak

Besides, several samples are prepared by following a different procedure in order to stop one or two destabilization mechanisms. To prevent ripening, we use perfluorohexane saturated nitrogen instead of pure nitrogen~\cite{2013_Cantat, 2017_Bey}. Chemical formula of perfluorohexane is $C_6F_{14}$ and it has a very low solubility in water, which reduces the global gas transfer rate between the bubbles. In addition, the effects of drainage were significantly reduced by making the samples rotate at 10~rpm around a horizontal axis for several hours after preparation.

\subsubsection{Stability of aqueous foam}

To check the ability of surfactant to stabilize the foam films during several hours, we prepare an initial height h$_0$=11~cm of aqueous foam in a glass column with diameter 2.6~cm. The generation method is the same as described in part~\ref{part_precursor_foam}, with initial bubble size $R_0 \simeq$ 300 $\mu$m. Foams with and without $C_6F_{14}$ were tested. At time t=0, wetting of the foam by imbibition is stopped (see paragraph \ref{part_precursor_foam} for details) and we record the evolution of the height $h(t)$ of the foam for at least one day. Both surfactants (TTAB at concentration 5~g/L and Steol at 1~g/L) were tested both in distilled water and in a synthetic cement pore solution containing 1.72~g/L of $CaSO_4\cdot2H_2O$, 6.959~g/L of $Na_2SO_4$, 4.757~g/L of $K_2SO_4$ and 7.12~g/L of $KOH$~\cite{2015_Bessaies}.

Note that the presence of perfluorohexane tends to make the foam swell. Indeed, there is $C_6F_{14}$ in the bubbles but not in the air in the column above the foam. Difference in $C_6F_{14}$ chemical potentials leads to transfer of nitrogen and oxygen from the air above the foam to the top bubbles. This swelling effect, which intrinsically results in a slight increase of foam height $h$, has been disregarded in the reported values of $h$.

\subsubsection{X-ray tomography}

Several cement foams have been studied by X-ray tomography one or two months after the preparation. Two types of experiments have been performed.

\begin{figure}[!ht]
\begin{center}
\includegraphics[width=5cm]{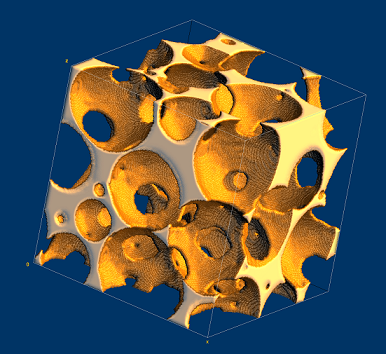}

\includegraphics[width=7cm]{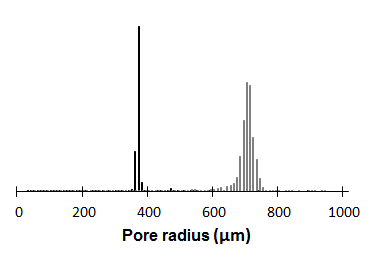}
\caption{Top: 3D reconstruction from X-ray tomography experiment on a Steol sample ($\tau_{y,0}=3$~Pa, $R_0$ = 685~$\mu$m). Cube size is 3~mm. Bottom: Final pore size distribution by volume (arbitrary unit on vertical axis) of the hardened cement foams, statistics on a 1.5~cm cube in the middle of the sample (about 2000 bubbles); black histogram illustrates a TTAB sample where W/C$_f$=0.42 and $R_0$=370 $\mu$m, and grey histrogram shows a Steol sample where $\tau_{y,0}=3$~Pa and $R_0$ = 685 $\mu$m. }
\label{picture_3D_R685}
\end{center}
\end{figure}

Images of one whole 2.6-cm-diameter and 6-cm-high sample and an additional 11-cm-high sample were obtained with a Ultratom scanner from RX solutions at Laboratoire Navier. Measurement involved a Hamamatsu L10801 X-ray source (160 kV) and a Paxscan Varian 2520V flat-panel imager. 
All scans were performed at 80~kV and 70~$\mu$A. To analyze the whole sample height, we have used stack type scans, i.e. horizontal sections of the sample were scanned independently and combined later by the reconstruction software. Frame rate was 3 images per second and 12 images were averaged to produce one projection. Resulting effective exposure time was therefore 4s.
3D tomographic reconstruction were performed at laboratoire Navier with the X-Act commercial software developed by RX-Solutions. Voxel size for the obtained images was 16.3~$\mu$m. Pores appeared black on the reconstructed images and interstitial cement pate was light grey. This allowed us to analyze the images with the freeware ImageJ program \cite{2012_Schneider} to compute the gaz volume fraction of the sample and the pore size distribution using the following procedure: first, a closing filter from MorphoLibJ plugin \cite{2016_Legland} with a 5 voxel-radius ball element was applied to reduce noise from the images. Then, image threshold was calculated using the Otsu method \cite{1979_Otsu}. On the one hand, gas volume fraction (see Fig. \ref{pictures_variation_air_fraction}) was deduced from the number of black pixels in each binarized horizontal slice. The binary images were also used to obtain the 3D visualization of the sample shown in Fig. \ref{picture_3D_R685}, top, with the 3D Viewer plugin. On the other hand, further processing was required to obtain the pore size distribution. 3D Watershed from MorphoLibJ was applied, then 1-voxel dilatation filter. Finally, 3D Object Counter plugin \cite{2006_Bolte} returned the volume $V_p$ of the pores. The obtained equivalent pore radii shown in Fig. \ref{picture_3D_R685}, bottom, were deduced from $V_p$ under the assumption of spherical pore shape. "Exclude on Edge" option prevented the incomplete bubbles on the image edge to be taken into account. Histograms in Fig. \ref{picture_3D_R685} confirms that that the pore size distributions obtained for two samples are narrow around the initial bubble radius, which validates our preparation protocol.

The microstructures of the elaborated materials have been characterized by means of synchrotron X-Ray microtomography on the recently constructed Anatomix beamline \cite{2017_Weitkamp} of the Soleil synchrotron facility, located in Saint Aubin, France, where we could add our sample during the measurement time of the proposal 20171213. Anatomix is a 200~m-long beamline based on a cryogenic in-vacuum undulateur (U18). The tomography setup was therefore still in a temporary state and was composed of a small air-bearing rotation table from LAB (RT150v3) equipped with a small manual goniometer head from X-Huber, which maintained a 20~mm long small PMMA rod at the end of which the sample, approximately 5 mm in diameter, was glued. X-Ray radiographs of the latter were obtained with a 20~$\mu$m-thick LuAG scintillator mounted on a right angle kinematic cage from Thorlabs, containing a mirror reflecting the optical image towards the CMOS sensor of a ORCA flash 4.0 v2 camera from Hamamatsu, through a 10x microscopic objective from Mitutoyo, with a numerical aperture of 0.28 and a working distance of 33.5~mm.
The optical definition of the sensor is 2048$\times$2048 pixels and the pixel size is 6.5~$\mu$m, inducing a voxel size of the tomographic reconstructions, assuming a perfect parallel projection, of 650~nm.
The undulator was set with a 7~mm gap and the beam was filtered by both a 10~$\mu$m Au filter and a 200~$\mu$m Cu filter, resulting in a large band pink beam centered on about 30~keV. Images were recorded synchronously with the continuous rotation of the table, at a frequency of 1~Hz, so that 2000 radiographs were recorded over a 180$\degree$ total angle, in about 35 minutes. In addition, 50 so-called "dark" images were recorded in the same conditions without the beam and 50 so-called reference images were recorded in presence of the beam but without the sample. The averages of these darks and references allowed us to compute the attenuation of the sample, assuming classically an affine dependance of grey levels with attenuation.
Finally, 3D tomographic reconstruction were performed at laboratoire Navier with the X-Act commercial software developed by RX-Solutions.
Obtained volumes were images of 2048$^3$ voxels, coded on 16-bits, providing 3D attenuations maps over a 1.33$\times$1.33$\times$1.33 mm$^3$ subvolume of the samples.
Even if their voxel size is of 650~nm, their actual spatial resolution is expected to be closer to 1~$\mu$m in their central part. Because of some residual imperfections of the manual alignment of the scintillator and the optics, images were slightly unfocused on their lower and upper parts, but the central zone of about 2048$\times$2048$\times$1200 voxels were of excellent quality.

\subsection{Properties of the reference cement pastes}

\subsubsection{Yield stress}
\label{part_reference_YS}

We estimate the yield stress of the reference cement pastes, i.e. prepared following the protocol described in~\ref{part_protocol_c5} with addition of foaming solution (without bubbles) instead of foam. Yield stress is measured using simple spread tests: the paste is poured on a flat horizontal surface and the yield stress is obtained by the following formula~\cite{2005_Roussel}:

\begin{equation}
\tau_{y,0} = \dfrac{225 \rho g \Omega^2}{128 \pi^2 R_{spread}^5}
\label{equation_seuil_c5}
\end{equation}
where $\rho$, $\Omega$ and $R_{spread}$ are respectively the density, the volume and the average radius of the spread cement paste. Note that this formula requires that $1~Pa\lesssim\tau_{y,0}\lesssim 100~Pa$~\cite{2005_Roussel,2017_Feneuil}. 

\bigbreak

\begin{figure}[!ht]
\begin{center}
\includegraphics[width=6cm]{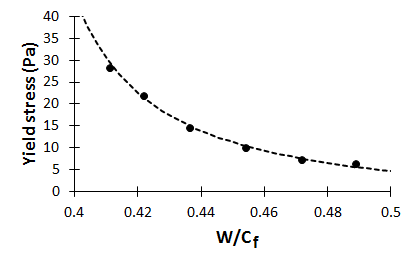}
\caption{Yield stress of cement pastes containing TTAB cationic surfactant and C1 cement (Fit with Yodel, equation~\ref{equation_Yodel_c5} with $m_1=15~Pa$, $\Phi_{perc}=0.32$ and $\Phi_{max}=0.46$)
}
\label{material_paste_yieldstress}
\end{center}
\end{figure}

TTAB cement foams are prepared with C1 cement. We have seen in a previous study~\cite{2017_Feneuil} that TTAB partially adsorbs on cement grains and has only a small effect on the yield stress. In such a case, the cement paste yield stress depends mainly on the final water-to-cement ratio W/C$_f$. Water-to-cement ratio of the cement paste before mixing with the precursor foam is W/C$_i$=0.37. Within our experimental conditions, both W/C$_f$ and final TTAB concentration in cement foam depend only on the liquid content in the precursor foam. Measured yield stress is given as a function of W/C$_f$ in Fig.~\ref{material_paste_yieldstress}. 

To get a relation between W/C$_f$ and $\tau_{y,0}$ ,which will allow us to calculate $\tau_{y,0}$ for all cement foams samples containing TTAB, we fit these data point using the so-called Yodel~\cite{2006_Flatt}. This model has been proposed to describe the yield stress of solid suspensions such as cement pastes. According to the Yodel, the yield stress can be obtained by the simple combination of a parameter $m_1$ which accounts for the interparticle forces, and a function of the solid volume fraction $\Phi_p$:

\begin{equation}
\tau_{y,0}=m_1 \dfrac{\Phi_p^2 (\Phi_p-\Phi_{perc})}{\Phi_{max} (\Phi_{max}-\Phi_p ) }
\label{equation_Yodel_c5}
\end{equation}
where $\Phi_{perc}$ is the percolation threshold and $\Phi_{max}$ is the maximal solid fraction. In cement pastes, $\Phi_p$ is related to the water-to-cement ratio: $\Phi_p=(\rho_w/\rho_c)/(\rho_w/\rho_c + W/C_f)$. 
Equation~\ref{equation_Yodel_c5} can be conveniently used for cement pastes containing TTAB because this surfactant hardly affects cement particle interactions, i.e. $m_1$ parameter is constant. 
The obtained fitting parameters are $m_1=15~Pa$, $\Phi_{perc}=0.32$ and $\Phi_{max}=0.46$.

\bigbreak

Steol cement foams are prepared with C2 cement. We have observed that Steol has a strong affinity with cement grain surface~\cite{2017_Feneuil}. Steol adsorption onto cement grains changes the interaction between the particles and modifies the yield stress of the cement paste. At low Steol concentration, yield stress increases due to Steol-induced hydrophobic interactions between cement grains. At high concentration, adsorbed Steol micelles create a steric repulsion between cement grains and strongly reduce yield stress. For the present study, we choose to use two Steol concentrations in the precursor cement paste: 11.4 and 12.4~g/L. In both cases, the addition of Steol into cement paste (step called "deflocculation" in Fig. \ref{schema_protocol_c5}) makes the yield stress drop to very low values, respectively 4~Pa and 1~Pa. Then, addition of Steol foaming solution at 1~g/L leads at the same time to an increase of W/C, which makes the yield stress decrease, and to a decrease of the Steol concentration, which makes the yield stress increase. Because of those two opposing effects on the yield stress, the latter is only weakly dependent on the small variations of liquid content in the precursor foam. Within our experimental conditions, after mixing with foam at liquid content 1.4\%, the final water-to-cement ratio is W/C$_f=$ 0.41. Steol concentration drops respectively to 10.4 and 11.4~g/L, and the yield stress increases to $\tau_{y,0} =$ 18~Pa and 3~Pa respectively.

\subsubsection{Surface tension of cement paste}

The global TTAB concentration after mixing of the precursor foam with the cement paste is between 0.7 and 2.4~g/L. Adsorption isotherms measured on pastes at W/C$_f$=0.5 showed that partial adsorption on cement grains leaves a residual concentration in solution between 0.2 and 1~g/L and corresponding surface tensions (in synthetic cement pore solution) are comprised between 37 and 42~mN/m~\cite{2017_Feneuil}. For the sake of simplicity we will assume that $\gamma_{TTAB}\simeq40~mN/m$. 

In the case of Steol surfactant, for both investigated concentrations, yield stress values are smaller than the yield stress of the same cement paste without surfactant (see paragraph \ref{part_reference_YS}). Therefore, all cement foams made with Steol are in the high Steol concentration regime, for which surface tension is $\gamma_{Steol}=27~mN/m$.

\section{Results}

\subsection{Stability of aqueous foams}

\begin{figure}[!ht]
\begin{center}
\includegraphics[width=6cm]{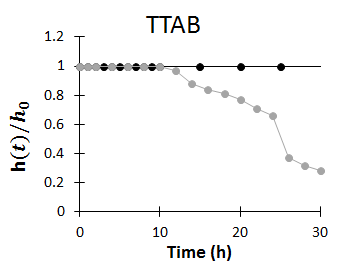}
\includegraphics[width=6cm]{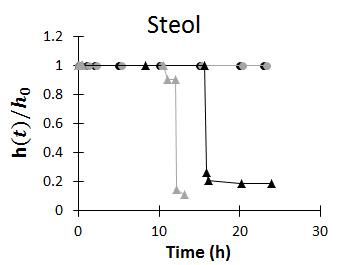}
\includegraphics[width=8cm]{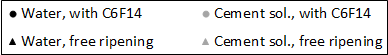}
\caption{Time evolution of the height of aqueous foams prepared from TTAB solutions at 5~g/L and Steol solutions at 1~g/L. Black dots or triangles refers to surfactant solutions in distilled water and grey signs to synthetic cement paste solutions. Dots refers to foams containing $C_6F_{14}$ to slow down ripening and triangles to foams made of nitrogen only.}
\label{graph_stability_aqueous}
\end{center}
\end{figure}

During the six experiments presented in Fig.~\ref{graph_stability_aqueous}, we observe that the foam becomes more and more dry and that the size of the air bubbles increases. However, the height of the foam did not decrease for 10~hours in all cases. Foams made with synthetic cement pore solution are less stable than those made from distilled water, and the presence of $C_6F_{14}$ decreases the collapse velocity; however, both these effects can be seen only after 10~hours.

\subsection{Drainage and ripening}

\subsubsection{Smaller bubbles ($R_0$ $\lesssim$ 500~$\mu$m)}

Let us first consider the unstable samples, whose properties are summarized in Table~\ref{Table_stability}. Preventing drainage by a rotation of the samples does not stop the foam destabilization process. On the other hand, counteracting artificially the ripening process using perfluorohexane allows for stabilizing robustly the fresh cement foams.

\begin{table}[!ht]
\begin{center}
\begin{tabular}{|l |c| c| c| c|}
\hline
$R_0$ & Free drainage & \textbf{No drainage} & \textbf{No ripening} &\textbf{No ripening} \\
($\mu$m) & and ripening & Free ripening & Free drainage & \textbf{No drainage} \\ \hline
\multicolumn{5}{|l|}{TTAB - C1, W/C$_f$ from 0.39 to 0.5}\\ \hline
$\approx$ 300 & Unstable & Unstable & STABLE & STABLE \\ \hline
$\approx$ 400 & Unstable & Unstable & STABLE & STABLE \\ \hline
\multicolumn{5}{|l|}{Steol - C2, $\tau_{y,0}=18$~Pa}\\ \hline
$\approx$ 200 & Unstable & Unstable & STABLE & STABLE \\ \hline
$\approx$ 300 & Unstable & Unstable & STABLE & STABLE \\ \hline
$\approx$ 400 & Unstable & Unstable & STABLE & STABLE \\ \hline
\multicolumn{5}{|l|}{Steol - C2, $\tau_{y,0}=3$~Pa}\\ \hline
$\approx$ 200 & Unstable & Unstable & STABLE & STABLE \\ \hline
\end{tabular}
\caption{Effect of slowing down ripening and drainage on sample stability.}
\label{Table_stability}
\end{center}
\end{table}

\begin{figure}[!ht]
\begin{center}
\includegraphics[width=4.5cm]{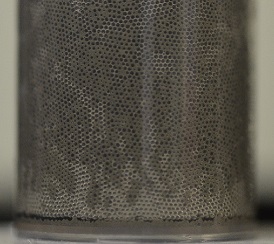}
\caption{Bottom of cement foam sample 15 minutes after production (TTAB, R = 300~$\mu$m, W/C$_f$=0.69 with perfluorohexane). Sample diameter is 2.6~cm.}
\label{pictures_drainage_noRip}
\end{center}
\end{figure}

When W/C$_f$ is further increased above 0.5 for TTAB samples (i.e. a yield stress deduced from equation 2 below 1 Pa) containing perfluorohexane, we sometimes notice a segregation of cement grains at the bottom of the samples, as illustrated in Fig.~\ref{pictures_drainage_noRip}. Bubble size after cement hardening in these samples is kept unchanged, except at the bottom of the sample. 

\begin{figure}[!ht]
\begin{center}
\begin{tabular}{l l l l}
\textbf{(a)} & \textbf{(b)} & \textbf{(c)} & \textbf{(d)}\\
\includegraphics[width=3cm]{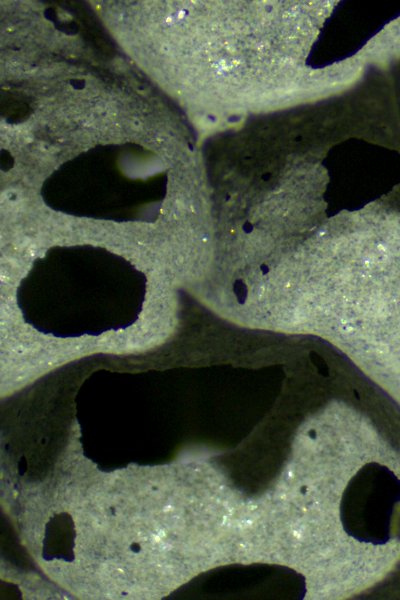} &
\includegraphics[width=3cm]{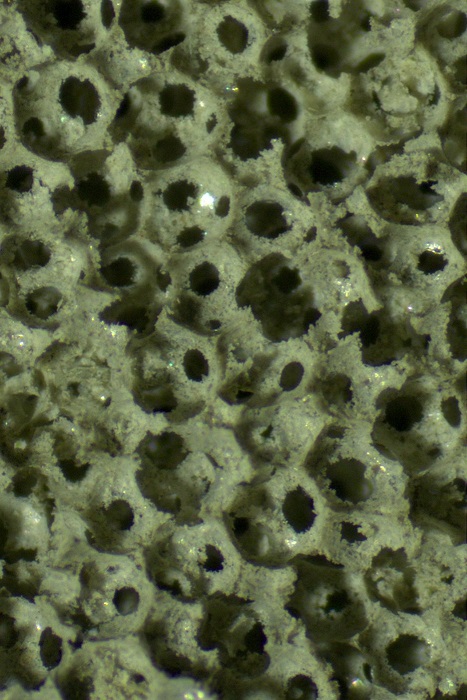} &
\includegraphics[width=3cm]{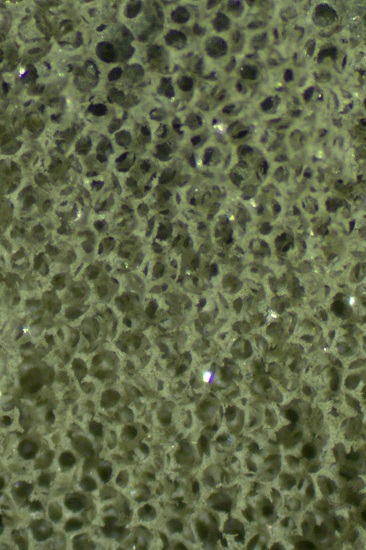} &
\includegraphics[width=3cm]{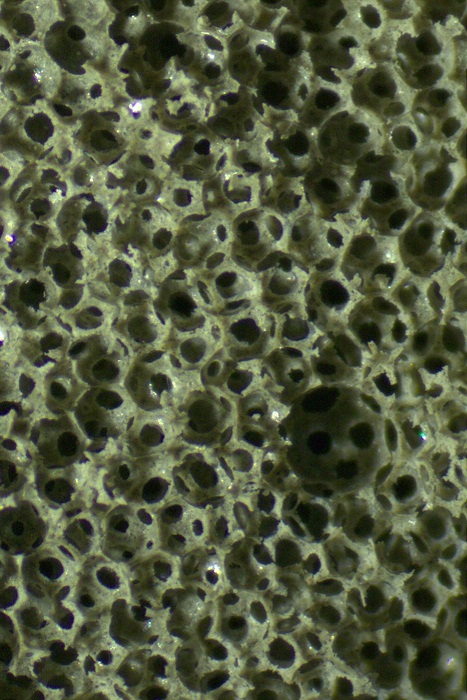} \\
\end{tabular}
\caption{Pictures of hardened cement foams. From left to right: (a) sample containing TTAB, initial bubble size 400~$\mu$m, with no control of destabilisation mechanisms; (b) sample containing Steol (18~Pa), initial bubble size 300~$\mu$m, with perfluorohexane; (c) example of inhomogeneous sample containing Steol (18~Pa), initial bubble size 200~$\mu$m, with perfluorohexane; (d) sample containing Steol (3~Pa), initial bubble size 200~$\mu$m, with perfluorohexane. Height for all pictures is 5~mm.}
\label{pictures_final_morphologies}
\end{center}
\end{figure}

Some pictures of the samples after cement hardening are shown as examples in Fig.~\ref{pictures_final_morphologies}. Each bubble is connected with its neighbors. Because air volume content is the same for all the samples (i.e. $\Phi$=83\%) the size of the opening between two bubbles depends only on the bubble size. Just after sample production and during several hours, the bubbles are separated by a liquid film containing no cement particle. When the samples are demolded, 7 days after sample preparation, the liquid film has already disappeared.

Note that for the smaller bubbles, mixing of cement paste and foam is sometimes not fully homogeneous at the bubble scale: small volumes of cement paste appear to be surrounded by foam of higher air content than the average value. For instance, structure of foams in pictures (c) and (d) in Fig.~\ref{pictures_final_morphologies} should be the same, but mixing was more inhomogeneous in sample (c).
However, the quality of the mixing does not change the fact that samples are stable only if ripening is prevented: the final pore distribution measured from the tomography images after cement hardening is narrow around the average value.

\begin{figure}[!ht]
\begin{center}
\includegraphics[width=0.24\textwidth]{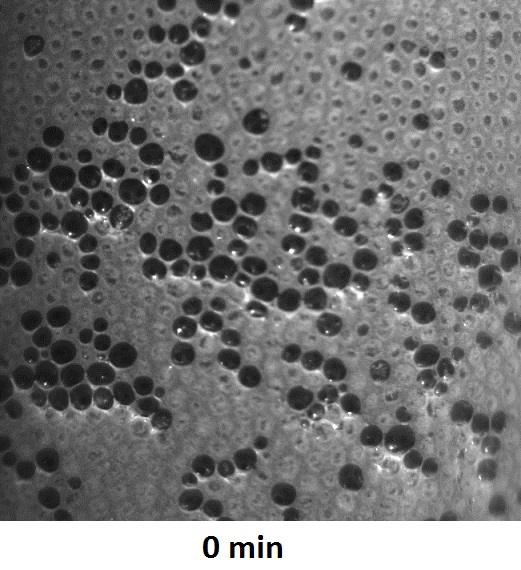}
\includegraphics[width=0.24\textwidth]{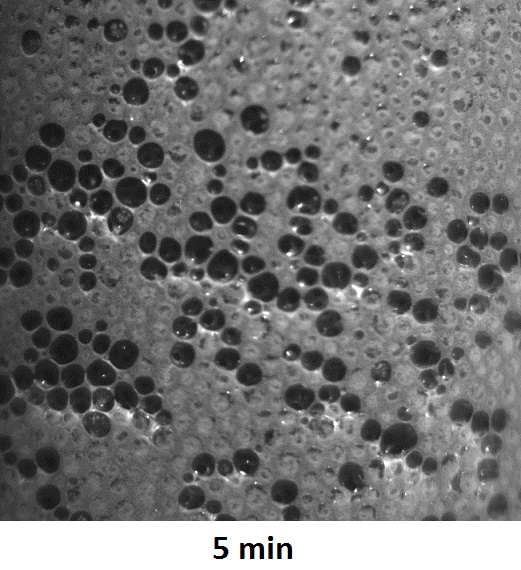}
\includegraphics[width=0.24\textwidth]{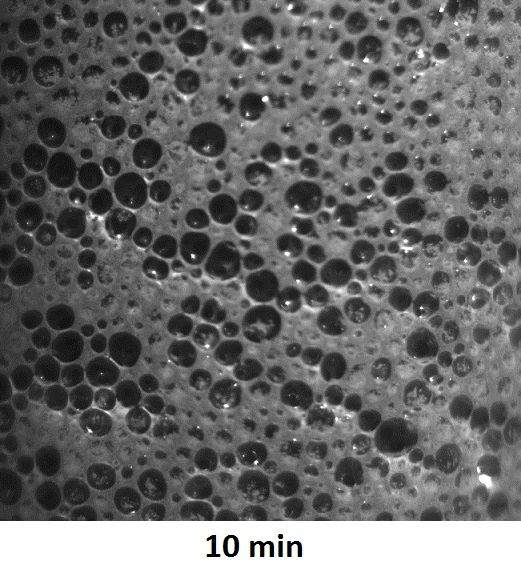}
\includegraphics[width=0.24\textwidth]{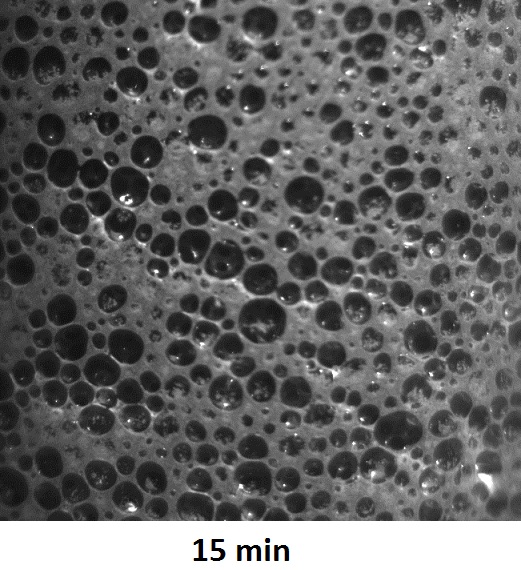}
\includegraphics[width=0.24\textwidth]{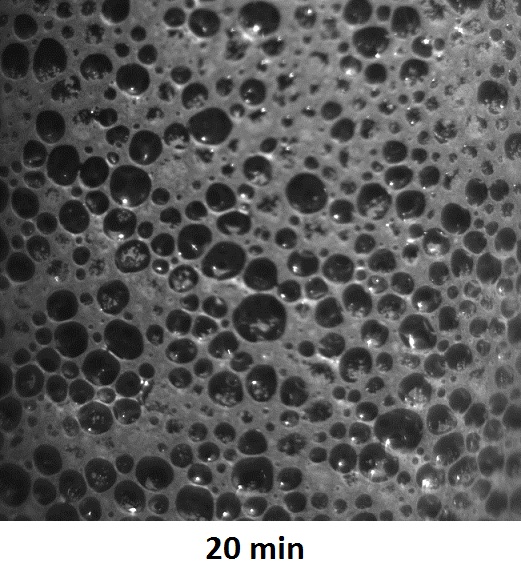}
\includegraphics[width=0.24\textwidth]{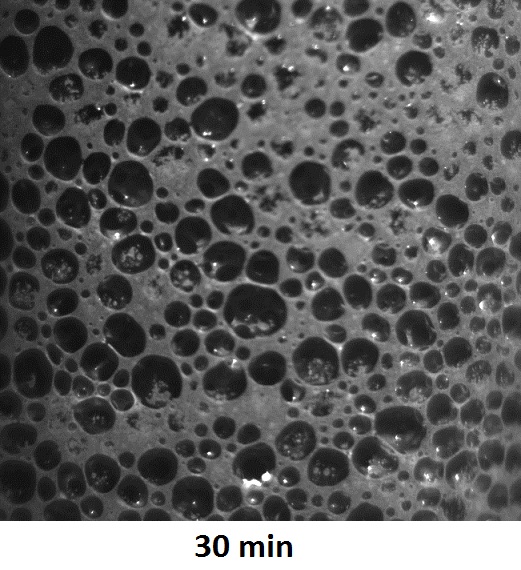}
\includegraphics[width=0.24\textwidth]{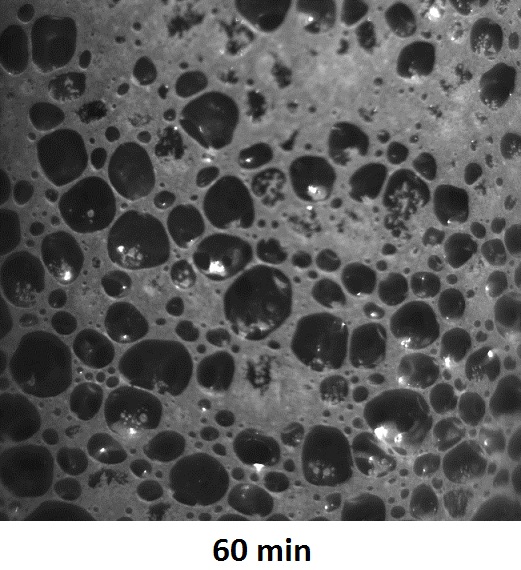}
\includegraphics[width=0.24\textwidth]{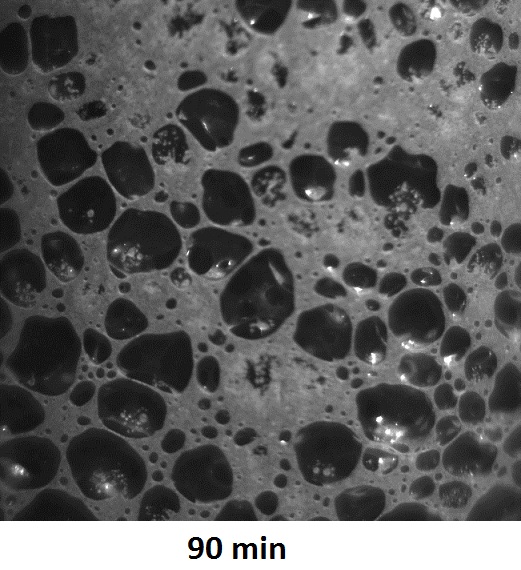}
\caption{Evolution of morphology of an unstable fresh cement foam. Example of sample with initial bubble radius 365~$\mu$m, W/C=0.41 and 83\% air content containing TTAB surfactant. Picture width is 1.5~cm.}
\label{images_ripening}
\end{center}
\end{figure}

\begin{figure}[!ht]
\begin{center}
\includegraphics[width=7cm]{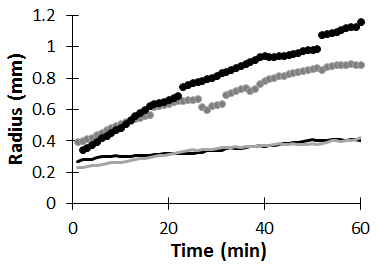}
\caption{Evolution of bubble radius with time for the TTAB samples with initial bubble radius 365~$\mu$m, W/C$_f$=0.41 (grey curves) and 270~$\mu$m, W/C$_f$=0.42 (black curves). Lines correspond to average value of all the bubbles and dots show the size of the defaults (average of the bigger three radius). Note that fast variations for the default size curves are due to the image analysis process: some of the bubbles cannot be identified on all the pictures because of unclean sample walls. In addition, the apparent radius are below the effective radius of the bulk bubbles due to the width of the Plateau borders on the pictures.}
\label{graph_ripening}
\end{center}
\end{figure}

An example for the evolution of the morphology of a TTAB cement foam with $R_0$ = 365~$\mu$m is shown in Fig.~\ref{images_ripening}. 
The average radius of the bubbles and the radius of the bigger bubbles (i.e. average of the bigger three radii) are given in Fig.~\ref{graph_ripening} for two TTAB samples with different initial bubble sizes but approximately the same W/C ratio (i.e. 0.41). For both samples, the maximal radius and the average radius of the bubbles start to increase just after sample preparation. Bubble size increases faster when the initial bubble size is smaller. More quantitatively, time for the bigger bubbles to reach twice their initial size is about 20 minutes after foam preparation when $R_0$ = 270~$\mu$m and 40 minutes after foam preparation when $R_0$ = 365~$\mu$m.

\subsubsection{Bigger bubbles ($R_0 \gtrsim$ 500~$\mu$m) }

\begin{figure}[!ht]
\begin{center}
\begin{tabular}{c c c}
h=3 cm & h= 10 cm &\\
\includegraphics[width=0.31\textwidth]{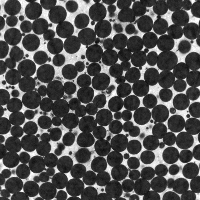} & 
\includegraphics[width=0.31\textwidth]{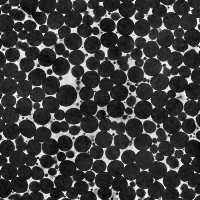} &
\includegraphics[width=0.31\textwidth]{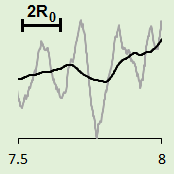}\\
\multicolumn{3}{c}{\includegraphics[width=0.95\textwidth]{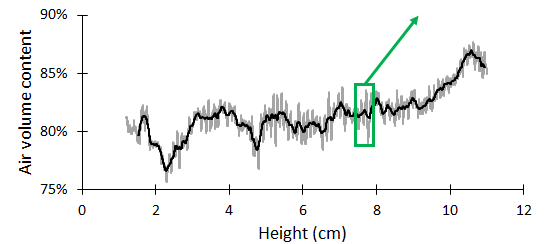}}
\end{tabular}
\caption{Steol sample ($\tau_{y,0}$=3~Pa, $R_0$=685~$\mu$m). Top: Slices of cement foam sample obtained by X-ray tomography at different height, bubbles appear in black and cement paste is grey. Bottom : evolution of gas fraction with sample height. Grey curve is raw data, black curve is the data average over 2000~$\mu$m.}
\label{pictures_variation_air_fraction}
\end{center}
\end{figure}

In samples containing larger bubbles, gas fraction often increases from the bottom to the top of the sample when drainage is not prevented through rotation. Examples of slices obtained by X-ray tomography of a 11~cm high sample are shown in Fig.~\ref{pictures_variation_air_fraction}. Gas fraction $\Phi$ has been calculated by image analysis of the tomography slices, and is also plotted in Fig.~\ref{pictures_variation_air_fraction} as a function of height: it increases from 80\% at the bottom to 85\% at the top. This curve additionally shows oscillations of $\Phi$ over a length scale close to 1300~$\mu$m, i.e. the diameter of the bubbles (see inset), which could be a signature of a bubble crystalline structure. Some local minima and maxima are still observed when the curve is averaged over a height bigger  than the bubble size (black curve). These local variations may be created during the mixing process or appear later during the drainage of the sample.

\bigbreak

\begin{figure}[!ht]
\begin{center}
\includegraphics[width=2.5cm]{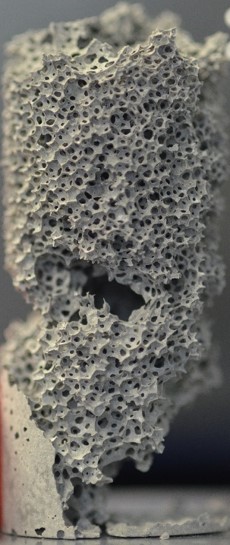}
\includegraphics[width=2.5cm]{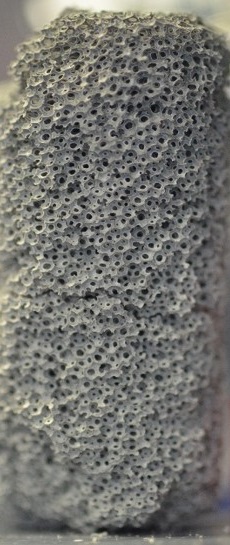}
\includegraphics[width=2.5cm]{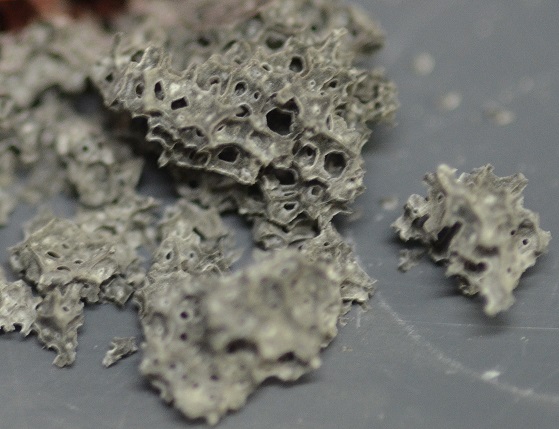}
\caption{Comparison of samples with big bubble size (TTAB, $\Phi$=0.87\%, $R_0$=1000~$\mu$m) after free drainage (W/C$_f$=0.44, left) and when drainage has been prevented by rotation (center and right). Rotation sometimes leads to weak samples that break when they are demolded (W/C$_f$=0.49, right). Sample height is 6~cm.}
\label{picture_drainage}
\end{center}
\end{figure}

In addition, drainage sometimes leads to a strong destabilization of bubbles at the bottom of the sample (see Fig.~\ref{picture_drainage}, left). Preventing drainage by sample rotation makes the sample more homogeneous (see Fig.~\ref{picture_drainage}, center) but often leads to weak samples, that break when they are demolded (see Fig.~\ref{picture_drainage}, right). This occurs when the cement foam is flowing in the mold during the rotation stage, which is often the case when $\tau_{y,0}$ is too low.

\newpage

\subsection{Influence of initial yield stress and bubbles size}

\begin{figure}[!ht]
\begin{center}
\includegraphics[width=12cm]{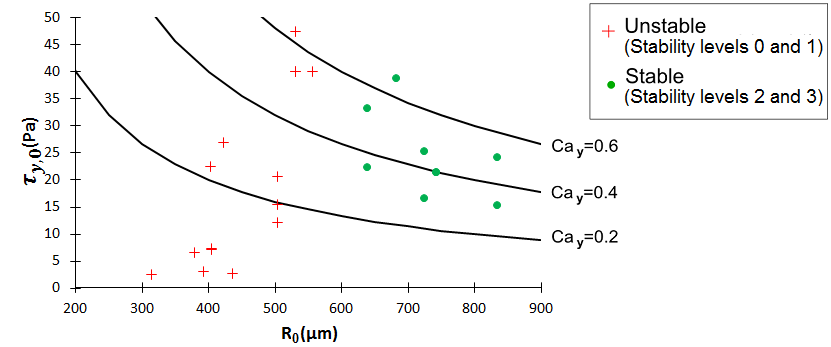}
\caption{Stability of TTAB cement foam samples. Black lines show constant Bingham capillary number  (as given by eq. \ref{equation_Bingham_Ca} with $\gamma_{TTAB}=~40mN/m$).}
\label{graph_stability_TTAB}
\end{center}
\end{figure}

\begin{figure}[!ht]
\begin{center}
\includegraphics[width=12cm]{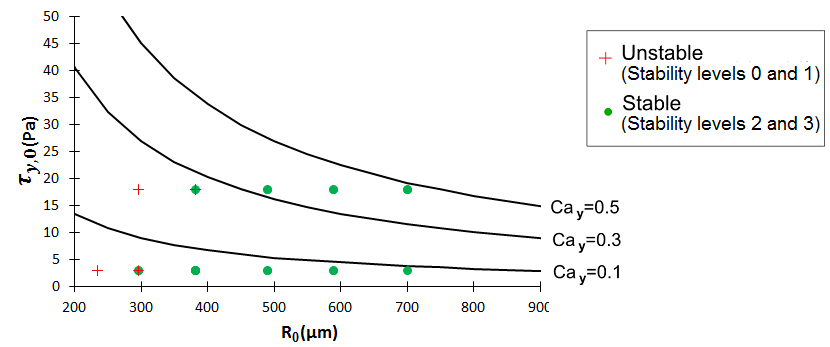}
\caption{Stability of Steol cement foam samples. Black lines show constant Bingham capillary number (as given by eq. \ref{equation_Bingham_Ca} with $\gamma_{Steol}=27~mN/m$).}
\label{graph_stability_Steol}
\end{center}
\end{figure}

In the following we do not consider the drainage-induced destabilization of the cement foam samples. A sample is considered as unstable when a major change of bubble size has occurred in the whole sample. Sample stability observations are gathered in Figs.~\ref{graph_stability_TTAB} and \ref{graph_stability_Steol}. For Steol, each point corresponds to at least two experiments. In some cases, identical foams have different stability behavior.

We observe for both surfactants a significant effect of bubble size: all samples are stable when $R_0$ is high and unstable at very low $R_0$. Minimum radius for stable foams depends however on the composition of the foams: 600~$\mu$m for TTAB and C1, and respectively 400~$\mu$m and 200~$\mu$m for Steol and C2 when $\tau_{y,0}$=18~Pa and 3~Pa. On the contrary, the increase of the yield stress does not improve stability in the studied range, i.e. below 50~Pa (we recall that higher yield stresses can not be achieved with our mixing device).

\newpage
\section{Discussion}

\subsection{Destabilization mechanisms}
\label{section_results_mechanism}

Let us first consider coalescence.  Experiments performed on aqueous foams made with both surfactants have shown that the foam volume was constant for more than 10 hours with $C_6F_{14}$ and without. Stability of a thin liquid film depends on the ability of the surfactant layers on both interfaces to repel each other. Film breakage occurs when the disjoining pressure $\Pi_d$, i.e. the pressure in the liquid film due the the repulsion on the air-liquid interfaces, reaches a critical value $\Pi_{d,crit}$. In an aqueous foam in mechanical equilibrium, the disjoining pressure is maximal at the top of the foam and $\Pi_d=\rho_{liq} g h$ where $\rho_{liq}$ is the liquid density. Our results show that TTAB and Steol are able to prevent coalescence, even in the highly alkaline conditions in cement paste, which means that  $\Pi_{d,crit} > 1000~Pa$. Note however that in foams, coalescence can take place even if $\Pi_{d}<\Pi_{d,crit}$, because bubble rearrangements can lead to film breakage due to dynamics, if liquid volume content is very low~\cite{2011_Biance}. Rearrangements can for instance be a consequence of ripening. This could explain why coalescence occurs after a few hours in perfluorohexane-free Steol foams whereas foams made with perfluorohexane were stable for 30~hours. These observations on aqueous foams show that coalescence is not expected to be the leading destabilisation mechanism in our cement foam samples. Let us now investigate the other two mechanisms. 

Drainage should stop if the yield stress of the cement paste exceeds a critical value $\tau_{c,d}$ whose order of magnitude is given by~\cite{2010_Guignot}:

\begin{equation}
\tau_{c,d} \sim \rho g r
\label{Equation_drainage}
\end{equation}
where $r$ refers to the radius of curvature of Plateau borders, as measured within their transversal cross-section. At high air content $\Phi>99\%$, $r \approx R_0/\sqrt{3(1-\Phi)}$~\cite{2013_Cantat} whereas for gas volume fractions close to the packing volume fraction of spherical bubbles (i.e. $\Phi\rightarrow 64\%$), $r\rightarrow R_0$. As $\Phi=83~\%$ in the present study, we take $r \sim R_0$, so $\tau_{c,d} \sim \rho g R_0$.

Ripening is expected to be slowed down or stopped when the yield stress of the interstitial material reaches the order of magnitude of the bubble capillary pressure~\cite{2014_Lesov}. Therefore, we define the critical yield stress for the ripening process as:

\begin{equation}
\tau_{c,r} \sim \gamma/R_0
\label{Equation_ripening}
\end{equation}

\begin{figure}[!ht]
\begin{center}
\includegraphics[width=9cm]{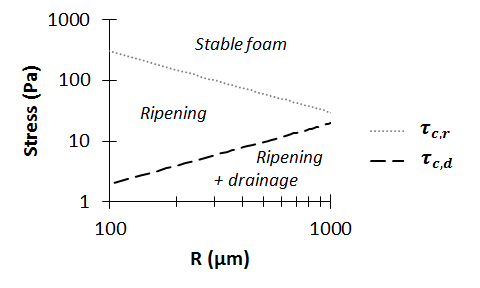}
\caption{Comparison of driving stresses for drainage and ripening. The curves define three possible behavior of cement foams when no experimental artifice is used to slow down ripening or drainage: stable foam if $\tau_{y,0}> \tau_{c,r}$, both drainage and ripening if $\tau_{y,0}< \tau_{c,d}$ and only ripening at intermediate yield stress values. }
\label{graph_drainageVSripening}
\end{center}
\end{figure}

Both critical stresses for drainage and ripening depend on the bubble radius $R_0$. Eqs \ref{Equation_drainage} and \ref{Equation_ripening} are plotted in Fig~\ref{graph_drainageVSripening} (we assume that $\rho\simeq 2000~$kg/m$^3$ and $\gamma \simeq 30$~mN/m). We notice that  $\tau_{c,d}<\tau_{c,r}$ within the range of studied bubbles sizes, i.e. 100~$\mu$m $<$ R $<$ 1000~$\mu$m. $\tau_{c,r}$ is two orders of magnitude above $\tau_{c,d}$ when R is a few 100~$\mu$m, which explains that ripening is the dominant destabilization mechanism at small bubble size. When $R \rightarrow$~1~mm, critical stresses for drainage and ripening are of the same order of magnitude. In this case, to understand the major destabilization mechanism at stake, we have to compare the kinetics of destabilization for drainage and ripening.

\subsection{Characteristic destabilization times}
\label{section_results_characteristic_times}

Let us first calculate a rough estimation of drainage characteristic time $t_d$ as given by the ratio of the sample height $H$ and the drainage velocity $v$: $t_d=H/v$~\citep{2013_Cantat}. The velocity is given by Darcy's law for pressure gradient induced by gravity $\rho$g:

\begin{equation}
v=\dfrac{k}{\mu_{app}} \rho g
\label{equation_Darcy_c5}
\end{equation}
where $k$ is the Darcy permeability and $\mu_{app}$ the apparent shear viscosity of the cement paste. Permeability of a foam depends of the gas volume fraction and the bubble size~\cite{2010_Rouyer}:

\begin{equation}
k = \dfrac{4(1-\Phi)^{3/2}}{1700(1-2.7(1-\Phi)+2.2(1-\Phi)^2)^2}R^2
\label{equation_permeability}
\end{equation}

For $\Phi=0.83$, this gives $k=5.10^{-4}R^2$.

To assess the apparent viscosity of the continuous phase, a rheological model must be chosen. The Bingham model describes well the rheological behavior of cement paste~\cite{2010_Roussel}, the stress $\tau$ and the shear rate $\dot{\gamma}$ in the paste are related by: 

\begin{equation}
\tau=\tau_{y} + \mu_p \dot{\gamma}
\label{Equation_BinghamModel}
\end{equation}
where the yield stress $\tau_{y}$ and the plastic viscosity $\mu_p$ depend on the paste formulation. We have measured the flow curves of three cement pastes with no surfactant, at water-to-cement ratio from 0.37 to 0.5 (results not shown here). For all three of them, stress at $\dot{\gamma}_{100}=100~s^{-1}$ is close to $2\tau_y$, i.e. $\mu_p\sim\tau_y/\dot{\gamma}_{100}$.

At the scale of the Plateau borders and nodes, the gravity-induced stress is $\tau \sim \rho g r$ where $r$ is the characteristic size of the channels and is close to the bubble radius $R_0$ as already explained in the previous paragraph. Following the above assumptions, the apparent viscosity is therefore:

\begin{equation}
\mu_{app}= \tau/\dot{\gamma}\sim\dfrac{\rho g R_0 \tau_y}{(\rho g R_0 - \tau_y)\dot{\gamma}_{100}}
\label{equation_apparentviscosity}
\end{equation}

From equations~\ref{equation_Darcy_c5}, \ref{equation_permeability} and \ref{equation_apparentviscosity}, we find that the characteristic drainage time is 

\begin{equation}
t_d \sim \dfrac{H \tau_y }{5.10^{-4}R_0 (\rho g R_0 - \tau_y)\dot{\gamma}_{100} }
\label{equation_drainage_time}
\end{equation}

This gives, for $\tau_y=1~Pa$, $t_d \sim 30~$min when $R_0$ = 200~$\mu$m and $t_d \sim 1~$min when $R_0$ = 700~$\mu$m.

\bigbreak

The characteristic time for foam ripening is given by~\cite{2012_Stevenson_Ripening}:

\begin{equation}
t_r=\dfrac{2 R_{av}^2}{K_2}
\label{equation_ripening_time}
\end{equation}
where $K_2$ is a diffusion coefficient. For nitrogen and low molecular weight surfactants $K_2 \sim 50~\mu m^2/s$. For polydisperse foams, the characteristic time given by equation \ref{equation_ripening_time} is the time during which the average bubble size grows from its initial value $<R>$ to 2 $<R>$. In the case of monodisperse foams, ripening is delayed: first, there is an induction period during which defaults appear and grow in the foam. The duration of this induction period is difficult to estimate. Here, we chose to use equation~\ref{equation_ripening_time} with $<R>=R_0$ to assess the time evolution of the defaults due to ripening.
We obtain $t_r \sim 50~$min when $R_0$ = 270~$\mu$m and $t_r \sim 90~$min when $R_0$ = 365~$\mu$m. These calculated times are of the same order of magnitude than the observed ripening times (see  Fig.~\ref{graph_ripening}).

\bigbreak

In table \ref{table_characteristic_times} we have reported both drainage and ripening characteristic times for several bubble sizes and several paste yield stresses. One can see that when cement paste yield stress is below the driving pressures for both drainage and ripening, drainage occurs faster than ripening as soon as $R_0 \gtrsim$ 200 $\mu$m. This result is compatible with the experimental observation that drainage sometimes occurs in samples containing the larger bubbles.

\begin{table}[!ht]
\begin{center}
\begin{tabular}{|r|c|c|c|}
\hline
 & R = 200~$\mu$m & 700~$\mu$m & 1~mm \\ \hline
$\tau_y = $ 1 Pa & Drainage + ripening $\sim$ 30~min & Drainage $\sim$ 1~min & Drainage $\sim$ 1~min \\ \hline
10 Pa & Ripening $\sim$ 30~min  & Drainage $\sim$ 1~h  & STABLE \\ \hline
100~Pa & Ripening $\sim$ 30~min & STABLE & STABLE \\ \hline
\end{tabular}
\caption{Characteristic destabilization times for cement foams with different bubble size and interstitial yield stress}
\label{table_characteristic_times}
\end{center}
\end{table}

\subsection{Effect of Bingham capillary number}

We have identified foam ripening as the major destabilization mechanism. To go further, we introduce the Bingham capillary number, which compares the yield stress (stabilizing effect) with the capillary pressure inside the bubbles, which drives the ripening process: 

\begin{equation}
Ca_y=\dfrac{\tau_y}{\gamma/R}
\label{equation_Bingham_Ca}
\end{equation}

\begin{figure}[!ht]
\begin{center}
\includegraphics[width=6cm]{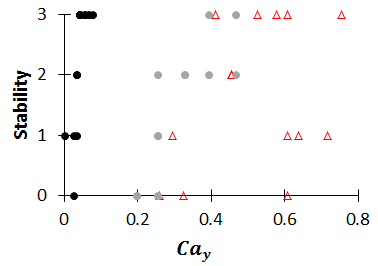}
\caption{Cement foam stability as a function of Bingham capillary number calculated from the yield stress of the reference cement paste. Black dots correspond to Steol samples with $\tau_{y,0}=3~$Pa, grey dots to Steol samples with $\tau_{y,0}=18~$Pa and empty triangles to TTAB samples.}
\label{graph_stability_CA_free}
\end{center}
\end{figure}

We expect that a suitable criterion for foam stability would be in the form of a critical capillary number $Ca_y^*$. As a first approach we use the yield stress of the reference cement paste $\tau_{y,0}$ (obtained from Fig.~\ref{material_paste_yieldstress}) to calculate the Bingham capillary number. The calculated values are plotted in Fig.~\ref{graph_stability_TTAB} and \ref{graph_stability_Steol} and compared to the stability of TTAB and Steol samples, showing that a critical value $Ca_y^*$ cannot be defined to describe the observed transitions between stable and unstable fresh cement foams. This can be better understood in Fig.~\ref{graph_stability_CA_free} where we have reported the observed stability for all samples as a function of the calculated $Ca_y$. One can see that stability is not controlled by the parameter $Ca_y$: in particular, the range of $Ca_y$ values for optimal stability depends on the surfactant used to make the cement foam.

\bigbreak

\bigbreak

$\tau_{y,0}$ is therefore not the adequate yield stress to be used to estimate the Bingham capillary number. Two effects should be considered:

\begin{itemize}
\item The yield stress of the interstitial cement paste, i.e. confined between the bubbles, can be higher than the yield stress of the reference cement paste, i.e. the unfoamed cement paste \cite{2019_Feneuil}. The interstitial yield stress $\tau_{y,int}$ is the effective stress that the bubbles must overcome to deform. It can be obtained from the macroscopic cement foam yield stress $\tau_{y,foam}(\Phi)$, as measured for this purpose, using the following equation~\cite{2017_Gorlier_4}:

\begin{equation}
\tau_{y,int}= \dfrac{\gamma/R_0}{c^{3/2} (1-\Phi)^2} \left( 1-\dfrac{\tau_{y,foam}(\Phi)}{\tau_{y,aq}(\Phi)} \right)^{3/2}
\label{equation_Gorlier_YieldStress_c5}
\end{equation}

where  $\tau_{y,aq}(\Phi)=0.6 \dfrac{\gamma}{R}(\Phi-0.64)^2$ is the yield stress of aqueous foam with the same $\Phi$ value and c=110 is a constant. 

\item Yield stress of cementitious materials at rest increases with time due to flocculation and creation of hydrate bonds between the particles~\cite{2010_Roussel}. In addition, it has been shown in a previous study~\cite{2019_Feneuil} that for some samples, densification of the interstitial yield stress increases during several minutes after the foam preparation due to liquid drainage and the associated densification of the cement paste. 
\end{itemize}

These two issues suggest that the interstitial yield stress $\tau_{y,int}$ at a relevant time $t^*$ after foam preparation should be considered instead of the reference yield stress $\tau_{y,0}$. The time $t^*$ is expected to be set by the interplay of several complex mechanisms, such as hydration, liquid drainage and foam ripening, and the detailed analysis for such effect is by far beyond the scope of this paper. We have noticed that all the unstable samples (including both samples presented in Figs. \ref{images_ripening} and \ref{graph_ripening}) have started to destabilize before 15 min. Therefore, in the following we consider that $t^*=$ 15~min, keeping in mind that this value could be different for cement foams prepared with a completely different protocol than ours. In the next part we measured the interstitial Bingham capillary number at $t^*=$ 15~min.

\subsection{15 min interstitial Bingham capillary number}

\begin{figure}[!ht]
\begin{center}
\includegraphics[width=6cm]{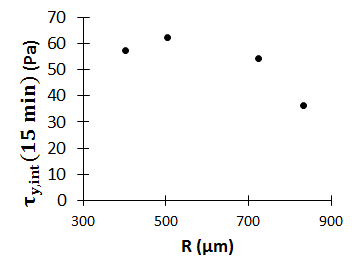}
\caption{Interstitial yield stress calculated from foam yield stress measured at 15~min for samples containing TTAB, W/C$_f$=0.42}
\label{graph_InterstialYS-R}
\end{center}
\end{figure}

We have measured the yield stress $\tau_{y,foam}$ of several cement foams at 15~min with the rheometry protocol described in \cite{2019_Feneuil}. The interstitial capillary number is then deduced from $\tau_{y,foam}$ using equation~\ref{equation_Gorlier_YieldStress_c5}. Results obtained for TTAB foams with different bubble sizes are presented in Fig. \ref{graph_InterstialYS-R}. The reported interstitial yield stress decreases for the largest bubble sizes. 

We stress that equation \ref{equation_Gorlier_YieldStress_c5} has been empirically obtained on model yield stress fluids (oil-in-water emulsion and beads suspension) within the range of $Ca_{y,int}$ values smaller than 0.5, whereas the present $Ca_{y,int}$ values reach unity for the largest bubble sizes. Therefore, it is possible that equation \ref{equation_Gorlier_YieldStress_c5} fails for $Ca_{y,int}$ approaching unity, and consequently the $\tau_{y,int}$ values obtained for the largest bubble sizes are not reliable.

\begin{figure}[!ht]
\begin{center}
\includegraphics[width=6cm]{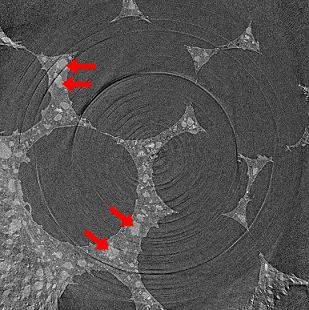}
\includegraphics[width=6cm]{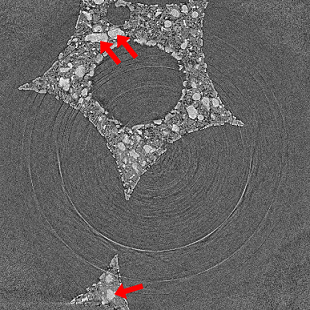}
\caption{Synchrotron tomography slices of TTAB cement foams stabilized by $C_6F_{14}$, for R=265~$\mu$m (left) and R=550~$\mu$m (right), two months after preparation. Image width is 1300~$\mu$m. Red arrows indicate some of the bigger cement grains.}
\label{pictures_synchrotron}
\end{center}
\end{figure}

On the other hand, a possible effect of bubble size is revealed by the high resolution tomography pictures in Fig.~\ref{pictures_synchrotron}. When the bubbles are small, the bigger cement grains, whose diameter before the start of hydration is about 100~$\mu$m, have similar size as the Plateau borders. Note that the synchrotron pictures have been taken two months after the sample preparation, when most of the cement has reacted with water and formed hydrates, resulting in a smaller apparent cement grain size than their initial size. The diameter of the larger particles that can enter the Plateau borders is given by equation~\cite{2010_Louvet}:

\begin{equation}
d_{PB}= 2 R \dfrac{0.27\sqrt{1-\Phi}+3.17(1-\Phi)^{2.75}}{1+0.57(1-\Phi)^{0.27}}
\label{equation_BdP_c5}
\end{equation}
with $\Phi=83\%$, we obtain $d_{PB}=60~\mu m$ if $R=300~\mu m$ and $d_{PB}=160~\mu m$ if $R=800~\mu m$. As a consequence, when the bubbles are small, larger cement grains cannot enter the Plateau borders and are stuck in the foam nodes. This is expected to lead to a segregation of cement grains according to their size, which could result in measurable effect on the yield stress of the cement paste.

\begin{figure}[!ht]
\begin{center}
\includegraphics[width=6cm]{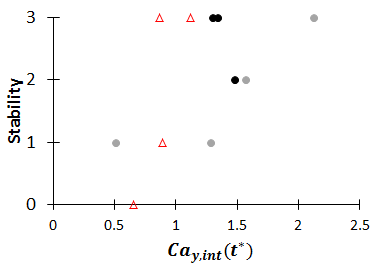}
\caption{Cement foam stability as a function of Bingham capillary number calculated from interstitial yield stress at 15~min. Black dots correspond to Steol samples with $\tau_{y,0}=3~$Pa, grey dots to Steol samples with $\tau_{y,0}=18~$Pa and empty triangles to TTAB samples.}
\label{graph_stability_CA_15min}
\end{center}
\end{figure}

Stability of samples is now plotted as a function the interstitial capillary number at $t^* = $ 15~min in Fig.~\ref{graph_stability_CA_15min}. We can see for all the curves that a transition from unstable samples at Ca$_{y,int}(t^*) \lesssim 1$ to stable samples when Ca$_{y,int}(t^*)$ is higher. We can therefore define a critical value for the Bingham capillary number: Ca$_{y,int}^*=1$. The stability criterion can therefore be expressed as follows:  

\textbf{\begin{equation}
Ca_{y,int}(t^*) > 1 
\end{equation}}


It is important to recall here the hypothesis leading to this criterion:
\begin{itemize}
\item Destabilization is due to ripening. We have shown in paragraphs \ref{section_results_mechanism} and \ref{section_results_characteristic_times} that this hypothesis is true for the samples studied here, but it may fail when bigger bubbles are studied ($R_0 \gtrsim$1 mm)
\item Interstitial yield stress has been calculated using equation \ref{equation_Gorlier_YieldStress_c5}. This empirical relation has been obtained for interstitial fluids that are continuous compared to the size of the Plateau borders, and for values of the Bingham capillary number below 0.5. Therefore, care should be taken when Plateau border size is small (due to small bubble size or high gas volume content) and when the interstitial yield stress is high.
\item The value of $t^*$ required to define this criterion is expected to depend on the preparation method of the cement foam samples and can be affected, for instance, by the polydispersity of the foam and the setting time of the cement. We are able to get a single value of $t^*$ for all the samples studied here because all are initially monodisperse, and although two different cements are used, hydration time for both is roughly similar. The obtained value, $t^*=15$ min is of the order of magnitude of the drainage characteristic time for the smallest studied bubbles (see paragraph \ref{section_results_characteristic_times}).
\end{itemize}

\section{Conclusion}

We have investigated the mechanisms at stake in the destabilization of cement foam samples. We first note that a proper choice of surfactant can prevent coalescence of the bubbles up to cement hardening. 

For most of the cement foam samples, prepared either with anionic or cationic surfactant, ripening is the leading destabilization mechanism. Drainage is also sometimes observed when the yield stress of cement paste is very low or bubble size is large. In these cases, we observed that drainage takes place faster than ripening, which was found to be consistent with our calculation of the drainage and ripening characteristic times.

Thanks to stability measurements performed on foams made from different formulations of cement pastes, i.e. surfactant and W/C ratio, with controlled bubble diameters, we have shown that the cement foam stability is governed by a single parameter, namely the Bingham capillary number calculated using the effective yield stress of the interstitial cement paste at time $t^*$ after the foam preparation, $Ca_{y,int}(t^*)$. Whatever the studied cement foam, the transition between stable and unstable foam is observed for the simple criterion $Ca_{y,int}^*(t^*) \simeq 1$.

In this work, the gas volume fraction of the samples has been kept constant and both cement used have similar setting times. In addition, we have studied foams with initially very narrow bubble size distribution. As setting time and bubble size distribution are expected to set the characteristic time $t^*$, it would be interesting to investigate how robust is our stability criterion when these parameters are allowed to vary within a large range.

\section*{Acknowledgments}

The authors wish to thank David Hautemayou and Cédric Mézières for technical support for the foam generation and mixing devices, Sabine Carré for support on synchrotron measurements, Nicolas Ducoulombier for help for the 3D reconstructions of tomography measurements, and Michel Bornert for giving us time to study our samples on the synchrotron line and for useful comments on this paper. This work has benefited from two French government Grants managed by the National Research Agency (ANR) [Grants no. ANR-11-LABX-022-01 and ANR-13-RMNP-0003-01]. In addition, the construction and operation of ANATOMIX is largely funded by the French ANR through the EQUIPEX investment program, project NanoimagesX, grant no. ANR-11-EQPX-0031.

\section*{References}
\bibliographystyle{elsarticle-num} 
\bibliography{Bibliographie}

\end{document}